\documentclass[prl,aps,reprint,superscriptaddress,showpacs,amsmath]{revtex4-1}

\usepackage{bm}
\usepackage{graphicx}
\usepackage{dsfont}

\usepackage{hyperref}
\usepackage{url}

\newcommand{\bra}[1]{\langle #1 | \,}
\newcommand{\ket}[1]{\, | #1 \rangle}

\newcommand{\expv}[1]{\langle #1 \rangle}

\newcommand{\Om}{\Omega}
\newcommand{\ga}{\gamma}
\newcommand{\Ga}{\Gamma}

\newcommand{\De}{\Delta}

\newcommand{\mc}[1]{\mathcal{#1}}
\newcommand{\sig}{\hat{\sigma}}

\begin{document}

\title{Steady-state crystallization of Rydberg excitations 
in an optically driven lattice gas}

\author{Michael H\"oning}
\affiliation{Fachbereich Physik und Forschungszentrum OPTIMAS, 
Technische Universit\"at Kaiserslautern, D-67663 Kaiserslautern, Germany}

\author{Dominik Muth}
\affiliation{Fachbereich Physik und Forschungszentrum OPTIMAS, 
Technische Universit\"at Kaiserslautern, D-67663 Kaiserslautern, Germany}

\author{David Petrosyan}
\affiliation{Institute of Electronic Structure and Laser, 
FORTH, GR-71110 Heraklion, Crete, Greece}

\author{Michael Fleischhauer}
\affiliation{Fachbereich Physik und Forschungszentrum OPTIMAS, 
Technische Universit\"at Kaiserslautern, D-67663 Kaiserslautern, Germany}

\date{\today}

\begin{abstract}
We study resonant optical excitations of atoms in a one-dimensional lattice
to the Rydberg states interacting via the van der Waals potential which 
suppresses simultaneous excitation of neighboring atoms. 
Considering two- and three-level excitation schemes, we analyze the dynamics 
and stationary state of the continuously-driven, dissipative many-body system
employing time-dependent density-matrix renormalization group (t-DMRG) 
simulations. We show that two-level atoms can exhibit only nearest 
neighbor correlations, while three-level atoms under dark-state resonant
driving can develop finite-range crystalline order of Rydberg excitations. 
We present an approximate rate equation model whose analytic solution 
yields qualitative understanding of the numerical results.
\end{abstract}


\maketitle

%%%% Introduction %%%%%%%%%%%%%%%%%%%%%%%%%%%%%%%%%%%%%%%%

Strong, long-range interaction between Rydberg atoms \cite{RydAtoms} 
have positioned them as promising systems for quantum information 
processing \cite{Jaksch2000,Lukin2001,rydrev}, which motivated considerable 
experimental progress in preparing and studying such systems  \cite{Tong2004,Singer2004,Vogt2006,Heidemann2007,Johnson2008,Reetz-Lamour2008,Urban2009,Gaetan2009}. Rydberg atoms are also interesting in the 
context of many-body physics: It was predicted that the long-range interaction
leads to spontaneous symmetry breaking and crystalline order in a continuous 
gas \cite{Weimer2008,Loew2009,Schwarzkopf2011}, formation of three-dimensional
super-solids \cite{Henkel2010}, as well as to vortex lattices \cite{Henkel2012}
or fractional quantum Hall states \cite{Grusdt2012} in two-dimensional (2D) 
systems with artificial magnetic fields. Coupling Rydberg atoms to light 
can give rise to novel photonic states with highly non-classical correlations 
\cite{Gorshkov2011,Peyronel2012} and unusual nonlinear spectroscopic features 
\cite{Pritchard2010,Petrosyan2011}.

For regular arrays of coherently driven atoms, depending on the strength 
and detuning of driving lasers, different ground-state phases with 
crystalline order emerge \cite{Schachenmayer2010,Lesanovsky2011,Lee2011}. 
Rydberg-dressed atoms \cite{Johnson2010,Pupillio2010,Hohner2010,Henkel2010}, 
i.e. ground-state atoms with a small admixture of Rydberg states, can also 
form crystalline structures with fractional fillings \cite{Lauer2012}.
In the optically driven lattice gas \cite{Viteau2011}, where the number 
of Rydberg excitations is not conserved, the preparation of the ground 
state of the system requires careful consideration and an adiabatic 
preparation scheme has been proposed and analyzed in \cite{Pohl2010}. 
A more natural approach to the formation of crystalline order of Rydberg 
excitations is to utilize the {\it stationary state} of a dissipative 
many-body system, which results from the interplay between continuous 
optical driving and spontaneous decay. Moreover, the steady state is 
an attractor of the system's dynamics and is therefore stable against 
small perturbations. 

Here we study resonant optical excitations of Rydberg states of atoms 
using two- and three-level driving schemes. The van der Waals (vdW) 
interaction between the atoms leads to a Rydberg level shift and thereby 
blocks the excitation of an atom which is sufficiently close to an already 
excited Rydberg atom. Specifically, we study a one-dimensional (1D) lattice 
with strong nearest neighbor interaction between the atoms, 
employing numerically exact t-DMRG simulations. We show that 
two-level driving leads to at most short-range correlations 
of Rydberg excitation probabilities of atoms at neighboring lattice sites. 
In contrast, for three-level atoms under the dark-state resonant driving, 
we find longer-range correlations and quasi-crystallization extending 
over several lattice periods. We derive an effective rate equations model, 
which can be solved exactly for the steady state yielding an expression 
for the correlation length. Finally, we estimate the influence of the tails 
of the vdW potential inducing small level shifts experienced by the next to 
nearest neighbor atoms, which give rise to an upper limit of 
the correlation length. 

%%%%%%%%%%%%%%%%%%%%%%%%%%%%%%%%%%%%%%%%%%%%%%%%%%
\begin{figure}[b]
\begin{center}
\includegraphics[width=0.9\columnwidth]{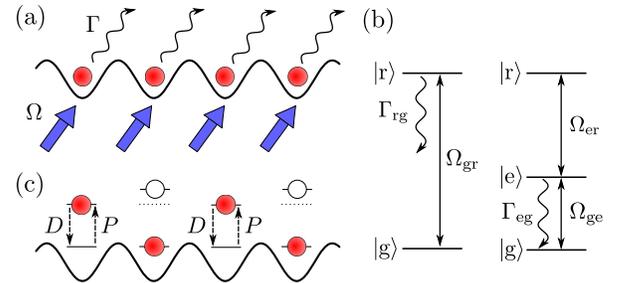}
\caption{(Color online).
(a) Schematics of optically driven atoms in a lattice.
(b) Starting from the ground state $\ket{g}$, the atoms are resonantly
excited to the Rydberg state $\ket{r}$ either directly (left)
or via resonant intermediate state $\ket{e}$ (right).
(c) Illustration of the rate equations model with Rydberg blockade 
of neighboring atoms.} 
\label{fig:scheme}
\end{center}
\end{figure}
%%%%%%%%%%%%%%%%%%%%%%%%%%%%%%%%%%%%%%%%%%%%%%%%%%%%

We thus consider a chain of $N$ atoms trapped in a 1D lattice potential 
of period $a$ [Fig.~\ref{fig:scheme}(a)] and examine
one- and two-photon resonant excitations of atoms from the ground 
state $\ket{g}$ to the Rydberg state $\ket{r}$ [Fig.~\ref{fig:scheme}(b)].  
In the two-level scheme, the transition $\ket{g} \to \ket{r}$ is 
driven by a laser field of (effective) Rabi frequency $\Om_{gr}$.
In the three-level scheme, the Rydberg state $\ket{r}$ is populated 
from the ground state $\ket{g}$ via resonant intermediate state 
$\ket{e}$ in the coherent population trapping (CPT) or dark-state
resonance configuration \cite{CPT-DTP_3la} with Rabi frequencies 
$\Om_{ge} \gtrsim \Om_{er}$. The corresponding atom-field 
interaction Hamiltonians are given, respectively, by 
$\mc{V}_{2}^j = -\hbar(\Om_{gr} \sig_{rg}^j + \mathrm{H. c.})$
and $\mc{V}_{3}^j = -\hbar(\Om_{ge} \sig_{eg}^j 
+ \Om_{er} \sig_{re}^j + \mathrm{H. c.})$, 
where $\sig_{\mu \nu}^j \equiv \ket{\mu}_{jj}\bra{\nu}$ 
are the transition operators for atom $j$. 

Spontaneous decay is described by Liouvillian terms 
in the equation of motion for the density operator $\rho$: 
$\mc{L}^j \rho = \frac{1}{2} \left( 2 \hat{L}^j \rho \hat{L}^{j\dagger} 
- \{ \hat{L}^{j \dagger} \hat{L}^{j},\rho \} \right)$,
where $\hat{L}^j$ are the Lindblad generators.  
For the two-level scheme, $\hat{L}_2^j = \sqrt{\Gamma_{rg}} \sig_{gr}^j$ 
with $\Ga_{rg}$ being the (population) decay rate of Rydberg state $\ket{r}$.
Although $\Ga_{rg}$ is typically small ($\sim 10^4\:$Hz), it may be  
comparable to the Rabi frequency $\Om_{gr}$ associated with either direct 
one-photon (UV) transition $\ket{g} \to \ket{r}$ with small dipole matrix 
element, or two-photon transition via far off-resonant intermediate states. 
For the three-level CPT scheme, we take into account only 
the large spontaneous decay rate $\Ga_{eg}$ ($\sim 10^7\:$Hz) 
of the intermediate exited state $\ket{e}$ via 
$\hat{L}_3^j = \sqrt{\Ga_{eg}} \sig_{ge}^j$; 
the decay rate $\Ga_{re}$ of Rydberg state can safely 
be neglected in comparison with $\Om_{er}$. 

For a single (isolated) two- or three-level atom under continuous
driving, the steady-state population of the Rydberg state is given, 
respectively, by
\begin{subequations}
\label{sigrr}
\begin{eqnarray}
\expv{\sig_{rr}} & \approx & \frac{|\Om_{gr}|^2}
{2 |\Om_{gr}|^2 + \ga_{rg}^2 + \De^2} , \label{sigrr2}  \\
\expv{\sig_{rr}} & \approx & \frac{|\Om_{ge}|^2 (|\Om_{ge}|^2 
+ |\Om_{er}|^2)} {(|\Om_{ge}|^2 + |\Om_{er}|^2)^2 
+ (\ga_{eg}^2 + 2 |\Om_{ge}|^2) \De^2}  , \qquad
\label{sigrr3} 
\end{eqnarray}
\end{subequations}
where $\ga_{\mu \nu} = \frac{1}{2} \Ga_{\mu \nu}$ and $\Delta$ is 
the one- or two-photon detuning of the laser frequency 
from the $\ket{g} \to \ket{r}$ transition resonance. 
It follows from Eqs.~(\ref{sigrr}) that the Rydberg state 
population $\expv{\sig_{rr}}$ of an atom is a Lorentzian 
function of $\De$, with the width 
$w = \sqrt{2 |\Om_{gr}|^2 + \ga_{rg}^2}$ for the direct excitation 
and $w = (|\Om_{ge}|^2 + |\Om_{er}|^2)/ \sqrt{\ga_{eg}^2 + 2 |\Om_{ge}|^2}$ 
for the CPT excitation schemes. 

Finally, pairs of atoms $i$ and $j$ interact with each other 
via the vdW potential \cite{rydcalc}
$\mc{V}_{\mathrm{vdW}}^{ij} = \hbar \sig_{rr}^i \frac{C_6}{d_{ij}^6} \sig_{rr}^j$,
where $d_{ij} = a|i-j|$ is the interatomic distance. Once an atom $i$ 
is excited to the Rydberg state $\ket{r}$, it shifts the atom $j$ 
out of the resonance by $\De = U/|i-j|^6$, where $U \equiv C_6/a^6$.
The excitation of atom $j$ to state $\ket{r}$ is then blocked if 
$\De \gtrsim w$, which determines the blockade distance 
$d_{\mathrm{b}} \simeq \sqrt[6]{C_6/w}$. Throughout this paper, 
we assume that $a< d_{\mathrm{b}} < 2a$, so that there is a 
Rydberg blockade only between the neighboring atoms.

The density matrix $\rho$ of the system of $N$ atoms obeys the master equation
\begin{equation}
\dot{\rho} = -\frac{i}{\hbar} [\mc{H}_l, \rho] +  \mc{L}_l \rho , \label{rhoME}
\end{equation}
with the Hamiltonian 
$\mc{H}_l = \sum_j \mc{V}_l^j + \sum_{i<j} \mc{V}_{\mathrm{vdW}}^{ij}$
and the Liouvillian $\mc{L}_l \rho =  \sum_j \mc{L}_l^j \rho$ 
for $l=2$ or $3$ level atoms. 

Since the interaction of an atom with its immediate neighbor is much 
stronger than that with the atoms further apart, we truncate the 
vdW potential $\mc{V}_{\mathrm{vdW}}^{ij}$ to nearest neighbor (NN),
$\mc{V}_{\mathrm{NN}}^{ij}= \hbar \sig_{rr}^i U \sig_{rr}^j$ for $i=j-1$
and $\mc{V}_{\mathrm{NN}}^{ij} = 0$ otherwise. 
We have performed exact numerical integrations of Eq.~(\ref{rhoME}) 
for small systems of several ($N \leq 7$) atoms interacting via 
the $\mc{V}_{\mathrm{vdW}}$ and $\mc{V}_{\mathrm{NN}}$ potentials, 
and verified that they yield similar results for both two- and
three-level excitation schemes. We will therefore 
employ from now on the $\mc{V}_{\mathrm{NN}}$ potential, 
returning to the corrections due to longer-range interactions 
of the $\mc{V}_{\mathrm{vdW}}$ potential later on.

We obtain the time evolution and the steady state of the full many-body 
density matrix Eq.~(\ref{rhoME}) employing the t-DMRG method 
\cite{Schollwoeck2011,Daley2004}. Our implementation follows the original 
proposal by Vidal \cite{Vidal2004}, generalized to open quantum systems 
\cite{Verstraete2004,Zwolak2004}. Simulations for up to $N \sim 10^2$ 
atoms are possible as the relaxation keeps the entanglement inside the 
system small. As a consequence, the usual limitations concerning the 
propagation time do not apply and the t-DMRG integration can be 
performed for, in principle, arbitrarily long times. Furthermore, 
since the stationary state is an attractor of the dynamics of 
the system, accumulated errors self-correct. We verified that 
for small systems the exact and t-DMRG solutions with the 
$\mc{V}_{\mathrm{NN}}$ potential are indistinguishable.

%%%%%%%%%%%%%%%%%%%%%%%%%%%%%%%%%%%%%%%%%%%%%%%%%%%%%%%%%
\begin{figure}[t]
\begin{center}
\includegraphics[width=0.9\columnwidth]{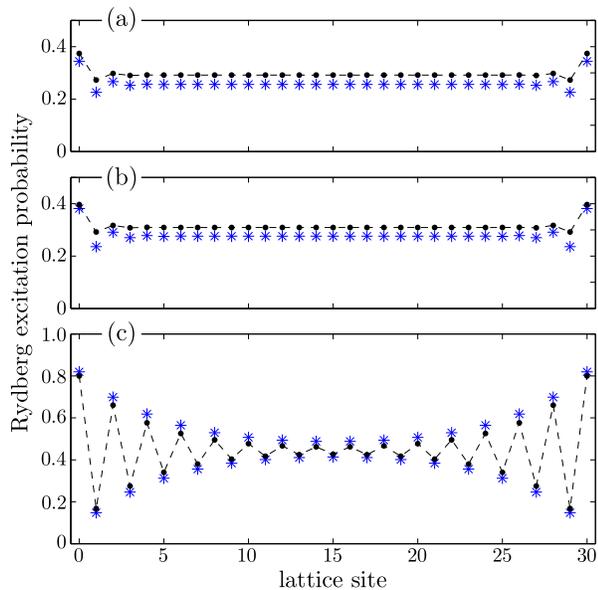}
\caption{(Color online). 
Steady-state Rydberg excitation probabilities $\expv{\sig_{rr}^j}$
of two- and three-level atoms in a lattice of length $30a$ ($N=31$ atoms).
Black dots (connected by dashed lines) show the t-DMRG solutions of 
Eq.~(\ref{rhoME}) with the truncated potential $\mc{V}_{\mathrm{NN}}^{ij}$,
and blue stars are the REs solutions for the hard-core (infinite NN) potential.
For the two-level atoms, the parameters are $U= 2 \Om_{gr}$, 
$\Ga_{rg} = \Om_{gr}$ (a) and $\Ga_{rg} = \frac{1}{4}\Om_{gr}$ (b).
For the three-level atoms, $U= 2 \Om_{ge}$, $\Ga_{eg} = 4 \Om_{ge}$,
and $\Om_{er} = \frac{1}{5}\Om_{ge}$ (c).}
\label{fig:steadyStateL31} 
\end{center}
\end{figure}
%%%%%%%%%%%%%%%%%%%%%%%%%%%%%%%%%%%%%%%%%%%%%%%%%%%%%%%%%

Results of the t-DMRG simulations for two- and three-level atoms in
a realistically large lattice of length $30a$ with open boundary 
conditions are shown in Fig.~\ref{fig:steadyStateL31}.  
Since interactions between the atoms suppress the Rydberg excitations, 
the atoms at the boundaries $j=0$ and $j=N-1$ having only one neighbor
acquire the largest population of the Rydberg state $\ket{r}$.
We also observe period-2 spatial oscillations of excitation 
probabilities $\expv{\sig_{rr}^j}$. For two-level atoms,
the amplitude of these oscillations, as well as the average 
Rydberg state population is low, 
both for weak $\Om_{gr} < \Ga_{rg}$ and moderately strong
$\Om_{gr} \simeq \Ga_{rg}$ driving, Fig.~\ref{fig:steadyStateL31}(a) and (b). 
This is due to the fact that even for a noninteracting two-level atom 
under strong driving $\Om_{gr} \gg \Ga_{rg}$ the excited state population 
saturates to $\expv{\sig_{rr}} \to \frac{1}{2}$ [cf. Eq.~(\ref{sigrr2})]. 
In contrast, for a (noninteracting) three-level atom under the CPT 
excitation scheme, the population of Rydberg state can be very 
large, $\expv{\sig_{rr}} \to 1$, when $\Om_{ge} \gg \Om_{er}$
[cf. Eq.~(\ref{sigrr3})]. 
For a chain of strongly interacting atoms, we then observe a large 
amplitude of spatial oscillations of the Rydberg excitation probabilities 
$\expv{\sig_{rr}^j}$ extending over many lattice sites, 
Fig.~\ref{fig:steadyStateL31}(c).

\begin{figure}[t]
\begin{center}
\includegraphics[width=0.9\columnwidth]{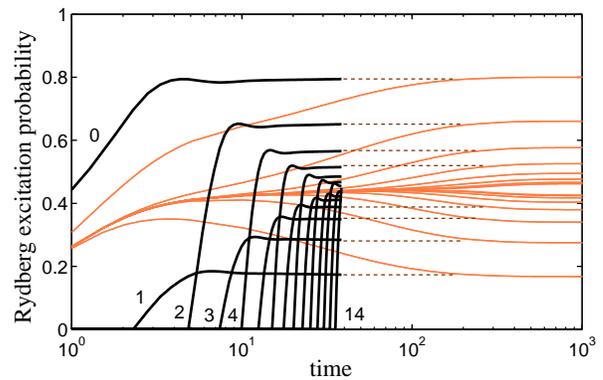}
\caption{(Color online).
Dynamics of Rydberg excitation probabilities
for simultaneous application of driving lasers 
at time $t=0$ to all the atoms (thin red lines), 
and for sequential application---``sweep''---of the lasers
to atoms $j=0,1,2,\ldots$ (thick black lines). 
All parameters are as in Fig.~\ref{fig:steadyStateL31}(c) 
and time is in units of $\frac{\Ga_{eg}}{\Om_{ge} \Om_{er}}$.}
\label{fig:dynamics_sweep}
\end{center}
\end{figure}

Atoms excited to Rydberg states are typically no longer trapped in 
an optical lattice, although there has been some progress in trapping 
Rydberg atoms using ponderomotive potentials \cite{Anderson2011}. 
An important question is therefore, whether the steady-state crystalline 
order of Rydberg excitations can be reached during times short enough 
for the center-of-mass motion of the excited atoms to be negligible.
In Fig.~\ref{fig:dynamics_sweep} we show the time evolution of Rydberg 
state populations of atoms initially in the ground state $\ket{g}$. 
Note that applying the driving lasers to all the atoms at once yields 
the global stationary state of the system after times two to three 
orders of magnitude longer than the steady-state equilibration time 
of an isolated atom. The reason for this is the initial formation of 
finite-size anti-ferromagnetic domains with dislocation defects 
in between, which require excessive healing times. But the steady-state
is independent of the initial conditions and details of preparation, 
therefore faster schemes for preparing the global stationary state of the 
system may exist. Similar to classical crystals, it is indeed much faster 
to ``grow'' the Rydberg quasi-crystal by first applying the lasers to 
the atoms at the lattice boundary and then gradually extending the 
irradiated region until the full Hamiltonian $\mc{H}_3$ is realized. 
In Fig.~\ref{fig:dynamics_sweep} we verify this intuitive approach. 
Since the crystalline order is rooted in the Rydberg blockade of 
the NN sites, a ``sweep'' velocity as fast as one lattice period $a$
per single-atom equilibration time $\Ga_{eg}/(\Om_{ge} \Om_{er})$ 
can be applied to reach the steady-state configuration.

%%%%%%%%%%%%%%%%%%%%%%%%%%%%%%%%%%%%%%%%%%%%%%%%%%%%%%%
%%%%%%%%%%%%%%%%%%%%%%%%%%%%%%%%%%%%%%%%%%%%%%%%%%%%%%%

We now describe an effective rate equations (REs) model, which yields an 
analytic solution for the stationary state of the system. We are interested
in the Rydberg excitation probabilities $\expv{\sig_{rr}^j}$ of atoms 
and their steady-state correlations $\expv{\sig_{rr}^j \sig_{rr}^k}$,
which allows us to disregard the coherences between the atoms. We restrict 
the interatomic interactions to the complete blockade of Rydberg excitations 
of NNs, assuming the potential $\mc{V}_{\mathrm{NN}}^{ij}$ with $U \to \infty$.
At each lattice site, we then have two incoherent processes: 
a state dependent pump with rate $P$ and a deexcitation with rate $D$, 
\begin{subequations}
\begin{eqnarray}
\hat{L}_{\mathrm{p}}^j &=& \sqrt{P} \, ( \sig^{j-1}_{rr} - \mathds{1} ) \, 
\sig^j_{rg} (\sig^{j+1}_{rr} - \mathds{1} ), \\
\hat{L}_{\mathrm{d}}^j &=& \sqrt{D} \, \sig^j_{gr} . 
\end{eqnarray}
\end{subequations}
The ratios of the two rates 
$\kappa_l \equiv \frac{P}{D} = \frac{1- \expv{\sig_{rr}} } {\expv{\sig_{rr}}}$ 
for the $l=2$ and $3$ level atoms are obtained from Eqs.~(\ref{sigrr}) as
\[
\kappa_2 = \frac{|\Omega_{gr}|^2}{|\Omega_{gr}|^2+\gamma_{rg}^2}, \qquad  
\kappa_3 = \frac{|\Omega_{ge}|^2}{|\Omega_{er}|^2}.
\]
In this model, the density matrix $\rho(t)$ obeys 
the equation of motion
\begin{equation}
\dot\rho = \sum_j \big(
  2 \hat{L}_{\mathrm{p}}^j \rho \hat{L}_{\mathrm{p}}^{j\dagger}
- \{\hat{L}_{\mathrm{p}}^{j\dagger}\hat{L}_{\mathrm{p}}^j,\rho \}
+ 2 \hat{L}_{\mathrm{d}}^j \rho \hat{L}_{\mathrm{d}}^{j\dagger} 
- \{\hat{L}_{\mathrm{d}}^{j\dagger}\hat{L}_{\mathrm{d}}^j,\rho \}\big).
\end{equation}
After sufficient relaxation time, the density matrix attains 
an essentially classical form 
$\rho(t)=\sum_{\{n_j\}} p(\{n_j\})(t)\,  \ket{\{n_j\}} \bra{\{n_j\}}$,
where $p(\{n_j\})$ is the probability of configuration 
${\{n_j\}\in (0,1)^N}$. Classical state-space dimension grows 
exponentially with $N$, precluding numeric integration for large systems.
However, the steady state of REs fulfills the detailed balance relation
\begin{equation}
\frac{p(\{m_j\})}{p(\{n_j\})}=\kappa_l^{\sum_{j}(m_j-n_j)}.
\label{eq:detailed-balance}
\end{equation}
States with the same number of excitations have equal weight 
and the partition function is given by 
$Z_N = \sum_{M=0}^{N} \Omega(M,N)\kappa_l^{M}$, where 
$\Omega(M,N)$ is the number of possible arrangements of $M$ excitations 
on a lattice of $N$ sites. In 1D, we have the analytic expression
$\Omega(M,N)=\binom{N-M+1}{N}$, which enables efficient calculation of all 
the steady state probabilities $p(\{n_j\})$.

In Fig.~\ref{fig:steadyStateL31} we compare the solutions of the REs model 
with the results of t-DMRG simulations, observing reasonable agreement, 
especially for the three-level excitation scheme of the atoms. In all cases, 
however, the decay of correlations is correctly captured by the REs model. 

For a large lattice $N \gg 1$, the partition function converges to 
$Z_N= \cosh \left( (2+N) \frac{\sqrt{\kappa_l}}{2}\right)$, with 
which the Rydberg excitation probabilities at odd sites $j$ 
(the boundary being at $j=0$) are given by
$ \expv{\sig_{rr}^j} = e^{-1/(2\sqrt{\kappa_l})}(1+e^{-j/\sqrt{\kappa_l}})/2$.
The decay of spacial oscillations of probabilities $\expv{\sig_{rr}^j}$
is characterized by the correlation length
\begin{equation}
\xi(\kappa_l)=\sqrt{\kappa_l} \, a . \label{eq:xi}
\end{equation}
Note that in a translationally invariant 
system (i.e. in the bulk), the excitation probability is uniform in space, 
but the onset of crystallization is revealed by the density-density 
correlations, which can also be calculated in the REs model. We find 
that $\xi(\kappa)$ is indeed the corresponding correlation length.   

For the two-level excitation scheme of atoms, $\kappa_2<1$, we then 
obtain the correlation length $\xi < a$, i.e. shorter than the lattice spacing. 
Using a mean-field approximation for the same system in the stationary state,
Lee {\it et al.} \cite{Lee2011} predicted a phase transition to 
anti-ferromagnetic order (period-2 density wave). 
The reason for this discrepancy is that the mean field approximation 
is inadequate in 1D. We note that even in a 2D square lattice the phase 
transition to the antiferromagnetic state cannot be reached with the 
two-level atoms; the 2D extension of the REs approach leads to the 
classical hard-square model, for which a phase transition to the 
N\'eel ordered (checkerboard) phase occurs at the critical value of 
$\kappa_{\textrm{crit}} \approx 3.7962$ \cite{Pearce1988}. 

For the CPT excitation of three-level atoms, one can tune $\kappa_3$ 
by the Rabi frequencies $\Omega_{ge}$ and $\Omega_{er}$, attaining large 
but finite correlation lengths $\xi > a$. Further increase of $\xi$ 
is constrained by the validity of the NN interaction approximation. 
We therefore consider the corrections originating from the hitherto 
neglected interactions of the atom with the next nearest neighbors (NNNs).
To estimate the maximal effect of the NNN interaction, let us assume 
a strong driving, resulting in Rydberg excitations at every other 
lattice site (half filling). A non-blocked atom is then likely to 
be detuned by $\De' = \De/2^6$ due to the interaction with the NNN 
Rydberg atoms. Expressing the NN detuning $\De = \beta w$ in terms 
of the excitation linewidth $w$ and using Eqs.~(\ref{sigrr3}), 
we obtain a modified
\begin{equation}
\kappa_3' =\frac{|\Omega_{ge}|^2}{|\Omega_{er}|^2 
+ (|\Omega_{ge}|^2 + |\Omega_{er}|^2) \left( \frac{\beta}{64} \right)^2}
\leq \left(\frac{64}{\beta}\right)^2. \label{eq:kappacorr}
\end{equation}
To ensure almost perfect NN blockade, it is reasonable to take 
$\beta \approx 10$ and $\sim 1\%$ excitation probability for the blocked atom.
In 1D, the correlation length is therefore limited to values of 
$\xi \lesssim 7 a$, due to the softness of the vdW interaction potential 
and the long wings of the Lorentzian excitation profile. 
In 2D, the phase transition to N\'eel order may occur for three-level atoms,
but since the effects of the NNN interactions will be more pronounced, 
the existence of a phase transition in 2D remains an open question. 
We finally note that for the dipole-dipole interaction 
$\mc{V}_{\mathrm{DD}}^{ij} \propto \frac{C_3}{d_{ij}^3}$, 
Eq.~(\ref{eq:kappacorr}) reduces to $\kappa_3' \leq (8/\beta)^2$ leading
to a short correlation length $\xi < a$.

To conclude, we have found that the collective steady-state of an ensemble of 
three-level atoms in a 1D lattice can exhibit quasi-crystallization of Rydberg 
excitations with a correlation length extending over many lattice periods. 
In contrast, for two-level atoms even under strong driving, the correlations 
are only between the neighboring atoms and the phase transition in 2D to N\'eel 
order cannot occur. Using the time-dependent DMRG 
simulations for several tens of atoms, we have shown that for uniform optical 
driving of all the atoms, long-range crystalline order of Rydberg excitations 
is attained only after very long times. A sequential excitation of 
neighboring atoms by dynamically ``sweeping'' the lattice with the driving 
laser, can result in the same steady-state of the system in a much shorter 
time given by the single-atom equilibration time multiplied by the number 
of sites. We derived an effective rate equations model whose exact 
steady-state solution, being in good agreement with the numerical calculations, 
yielded explicit analytic expressions for the Rydberg excitation probabilities 
and correlation length.

\begin{acknowledgments} 
Financial support of the Deutsche Forschungsgemeinschaft 
through SFB TR49 is acknowledged. 
D.P. is grateful to the University of Kaiserslautern 
for hospitality and support.
\end{acknowledgments}

\end{document}